\documentclass{appolb}
\usepackage{graphicx}

\begin{document}
\title{In-medium $\bar K$- and $\eta$-meson interactions and bound 
states\footnote{Presented by Avraham Gal at the II International Symposium 
on Mesic Nuclei, Cracow, September 22-24 2013.}} 
\author{Avraham Gal, Eli Friedman, Nir Barnea 
\address{Racah Institute of Physics, The Hebrew University \\ 
Jerusalem 91904, Israel} 
\and  
Ale\v{s} Ciepl\'{y}, Ji\v{r}\'{i} Mare\v{s}, Daniel Gazda\footnote{Present 
address: ECT*, Villa Tambosi, I-38123 Villazzano (Trento), Italy.}
\address{Nuclear Physics Institute, 25068 \v{R}e\v{z}, Czech Republic}} 

\maketitle

\begin{abstract}

The role played by subthreshold meson-baryon dynamics is demonstrated in 
$K^-$-atom, $\bar K$-nuclear and $\eta$-nuclear bound-state calculations 
within in-medium models of $\bar K N$ and $\eta N$ interactions. New 
analyses of kaonic atom data reveal appreciable multi-nucleon contributions. 
Calculations of $\eta$-nuclear bound states show, in particular, that the 
$\eta N$ scattering length is not a useful indicator of whether or not $\eta$ 
mesons bind in nuclei nor of the widths anticipated for such states. 

\end{abstract}

\section{Introduction} 
\label{sec1} 

The near-threshold $\bar K N$ and $\eta N$ scattering amplitudes are both 
attractive and energy dependent in models that generate dynamically the 
nearby meson-baryon $s$-wave resonances $\Lambda(1405)$ and $N^{\ast}(1535)$, 
respectively. Although free-space hadron-nucleon attraction at threshold 
appears to be a necessary condition for binding hadrons in nuclei, 
careful consideration of medium modifications is required to turn it into 
a sufficient condition. For example, it was pointed out by Wycech and others 
in the early 1970s that {\it subthreshold} $K^-N$ scattering amplitudes are 
the relevant ones even for studies of kaonic atoms, in spite of the kaon 
energy essentially being at threshold \cite{wycech71,BT72,rook75}. 
Yet, systematic treatments of energy dependence within dynamical and 
self-consistent calculations of $\bar K$ and $\eta$ bound states in nuclei 
have been lacking until recently. The present overview is focused on recent 
progress made by the Prague-Jerusalem Collaboration towards incorporating 
medium modifications, particularly those implied by the energy dependence 
of the corresponding scattering 
amplitudes \cite{CFGGM11,CFGK11,BGL12,FG12,GM12,FG13,FGM13,CFGM13}. 

\begin{figure}[htb] 
\begin{center} 
\includegraphics[width=1.0\textwidth]{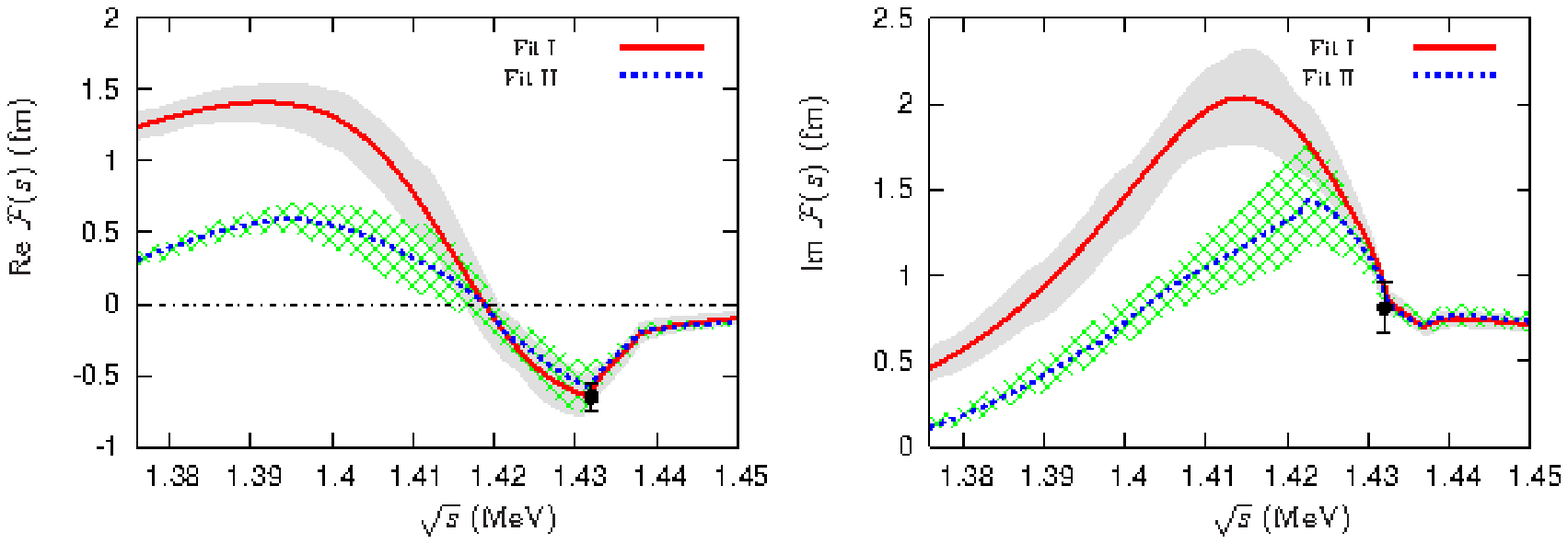} 
\caption{Real (left panel) and imaginary (right panel) parts of the $K^-p$ 
center-of-mass (cm) scattering amplitudes generated in two NLO chiral-model 
fits \cite{GO13}. The $K^-p$ threshold values marked by solid dots follow 
from the SIDDHARTA measurement of kaonic hydrogen $1s$ level shift and 
width \cite{SID11}. Figure adapted from Ref.~\cite{GO13}.} 
\label{fig:oller} 
\includegraphics[width=0.48\textwidth]{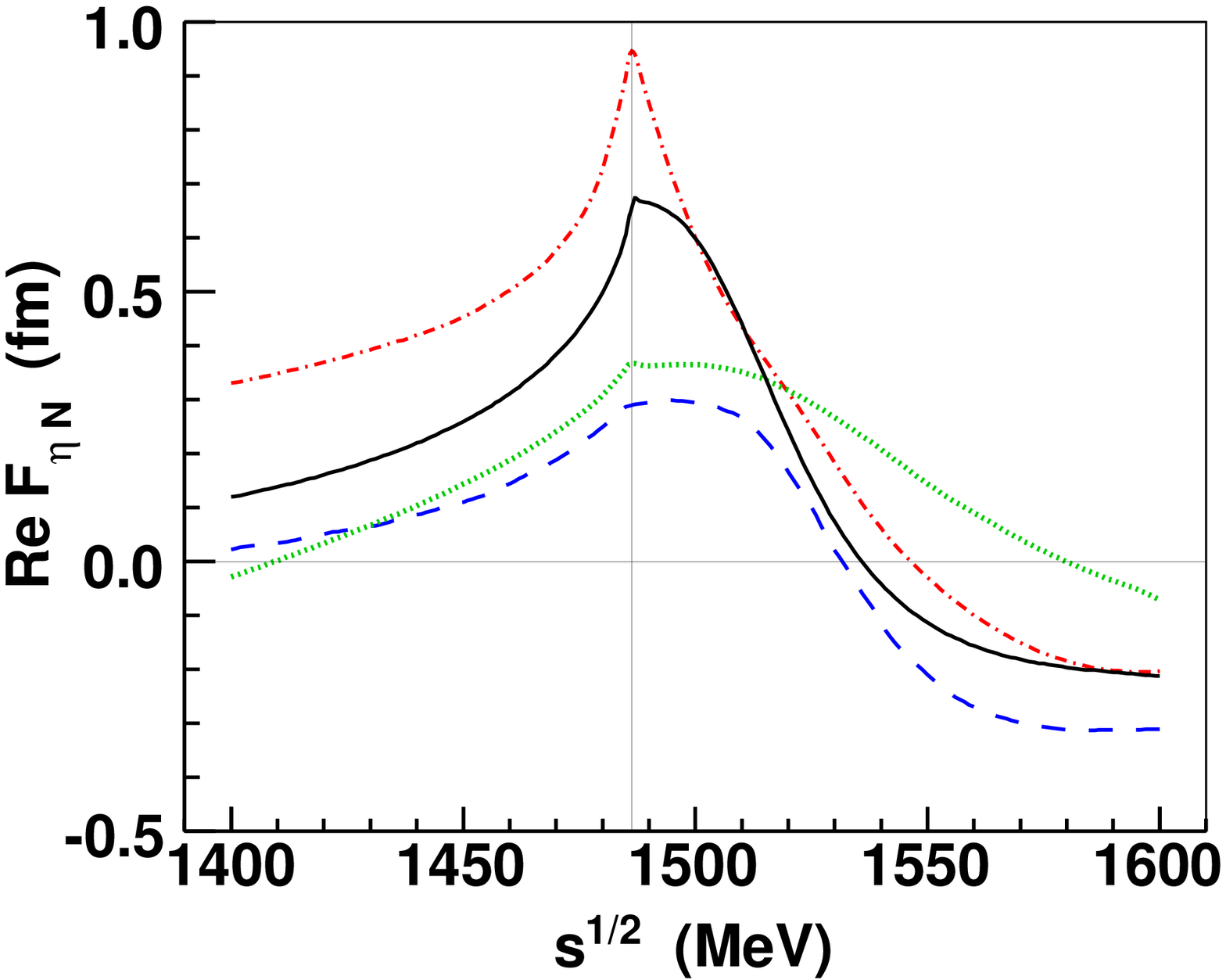} 
\includegraphics[width=0.48\textwidth]{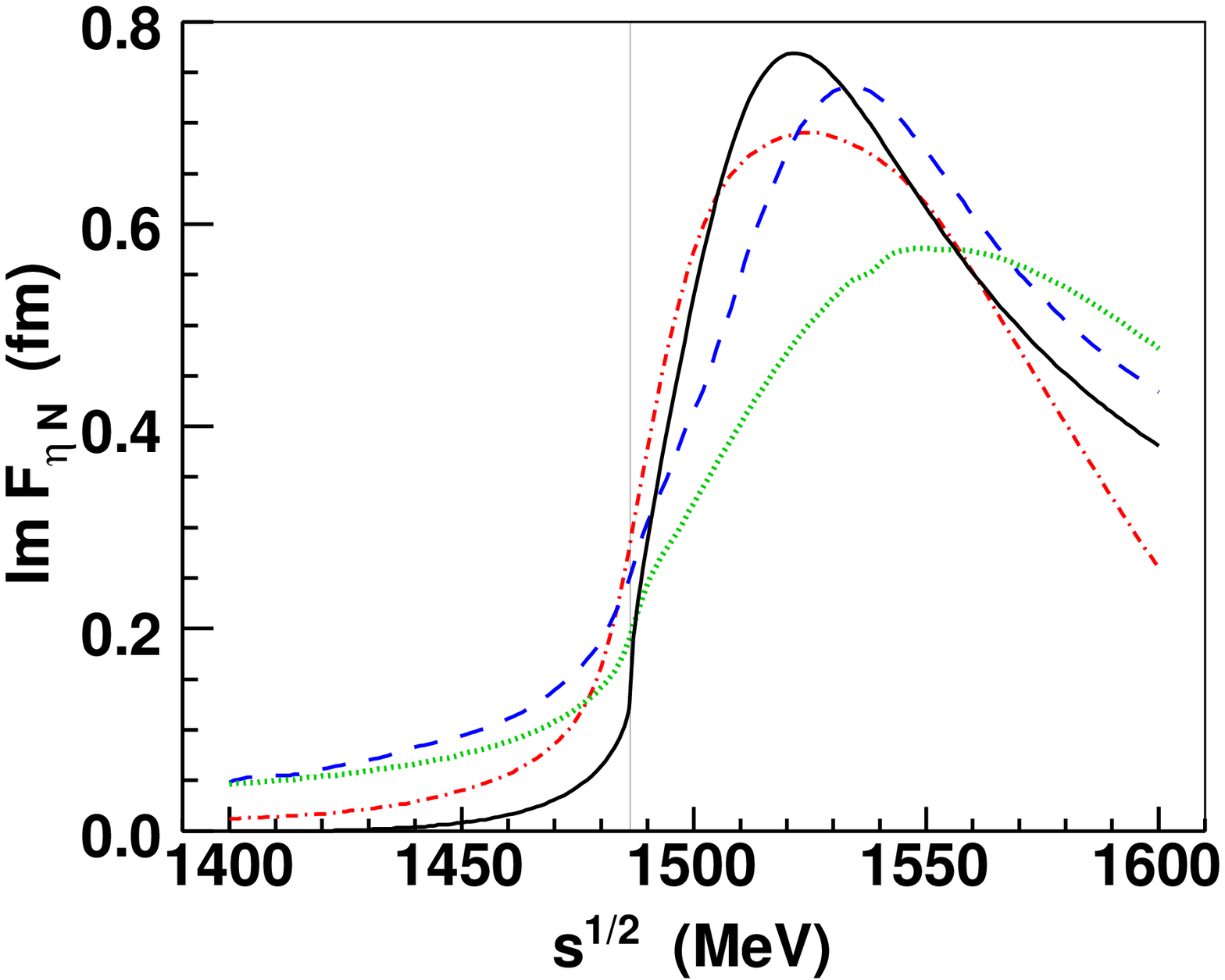} 
\caption{Real (left panel) and imaginary (right panel) parts of the $\eta N$ 
cm scattering amplitude $F_{\eta N}(\sqrt{s})$ as a function of the total cm 
energy $\sqrt{s}$ in four meson-baryon coupled-channel interaction models. 
In decreasing order of Re$\;a_{\eta N}$: dot-dashed, GW \cite{GW05}; solid, 
CS \cite{CS13}; dotted, M2 \cite{MBM12}; dashed, IOV \cite{IOV02}. 
The thin vertical line denotes the $\eta N$ 
threshold.} 
\label{fig:aEtaN1} 
\end{center} 
\end{figure} 

The first point worth noting about subthreshold scattering amplitudes is 
that they are highly model dependent. This is demonstrated for $K^-p$ in 
Fig.~\ref{fig:oller} and for $\eta N$ in Fig.~\ref{fig:aEtaN1}. For $K^-p$ 
the two NLO chiral-model fits by Guo and Oller \cite{GO13} to scattering and 
reaction data above and at threshold generate scattering amplitudes, shown 
in Fig.~\ref{fig:oller}, that differ substantially from each other in the 
subthreshold region. Fit-I amplitude is quite similar to those found in NLO 
fits by Ikeda, Hyodo and Weise (IHW) \cite{IHW11} and by Ciepl\'{y} and 
Smejkal (CS) \cite{CS12} both of which were used in recent $K^-$-atom 
calculations \cite{FG12,FG13}. For $\eta N$ the four scattering amplitudes 
shown in Fig.~\ref{fig:aEtaN1} differ below as well as above the $\eta N$ 
threshold, with perhaps just one common value $a_{\eta N}\approx 0.2-0.3$~fm 
for the imaginary part at threshold. 

The present overview is organized as follows. In-medium meson-baryon 
scattering amplitudes are discussed in Sect.~\ref{sec2}, for both $\bar KN$ 
and $\eta N$, focusing on the connection between their (subthreshold) 
energy and density dependencies. The use of such in-medium $K^-N$ 
scattering amplitudes in kaonic-atom calculations and fits is discussed in 
Sect.~\ref{sec3}. Related applications to kaonic bound-state calculations are 
discussed in Sect.~\ref{sec4} for few-body systems, and in Sect.~\ref{sec5} 
for many-body systems. For $\eta$ mesons, nuclear bound-state calculations 
using in-medium energy and density dependent $\eta N$ amplitudes are discussed 
in Sect.~\ref{sec6}. A brief summary and outlook in Sect.~\ref{sec7} concludes 
this presentation.

\section{In-medium amplitudes and energy versus density dependence}  
\label{sec2} 

The in-medium modifications of free-space scattering amplitudes become 
particularly transparent by working with separable interactions. In 
Refs.~\cite{CS13,CS12} Ciepl\'{y} and Smejkal introduced meson-baryon 
coupled-channel energy-dependent separable $s$-wave interactions matched to 
SU(3) chiral scattering amplitudes in up to next-to-leading order (NLO) of 
the chiral expansion. 
The Tomozawa-Weinberg leading order (LO) term provides a good approximation 
for $\bar{K}N$ \cite{CS12}, but going to NLO is mandatory for $\eta N$ 
\cite{CS13} since the relevant data involve dominantly the $\pi N$ channel 
which is decoupled from $\eta N$ at LO. Solving the in-medium coupled-channel 
Lippmann-Schwinger equations $F=V+VGF$ with these potential kernels leads to 
a separable form of in-medium scattering amplitudes $F_{ij}$, given in the 
two-body cm system by 
\begin{equation}
F_{ij}(k,k';\sqrt{s},\rho)=g_{i}(k^{2}) \: f_{ij}(\sqrt{s},\rho) \: 
g_{j}(k'^{2}) \; ,
\label{eq:Fsep}
\end{equation}
with momentum-space form factors $g_{j}(k^{2})$, where $j$ runs over channels, 
and in-medium reduced amplitudes $f_{ij}(\sqrt{s},\rho)$ expressed as 
\begin{equation}
f_{ij}(\sqrt{s}, \rho)=\left[ (1 - v(\sqrt{s}) \cdot G(\sqrt{s}, \rho))^{-1}
\cdot v(\sqrt{s}) \right]_{ij} \; .
\label{eq:fij}
\end{equation}
Here, $G$ is a channel-diagonal Green's function in the nuclear medium: 
\begin{equation}
G_{n}(\sqrt{s},\rho) = -4\pi \: \int_{\Omega_{n}(\rho)}
\frac{d^{3}p}{(2\pi)^{3}}\frac{g_{n}^{2}(p^{2})}
{k_{n}^{2}-p^{2} -\Pi^{(n)}(\sqrt{s},\rho) +{\rm i}0} \; ,
\label{eq:Grho}
\end{equation} 
where the integration on intermediate meson-baryon momenta is limited to 
a region $\Omega_{n}(\rho)$ ensuring that the intermediate nucleon energy 
is above the Fermi level in channels $n$ involving nucleons. The self-energy 
$\Pi^{(n)}(\sqrt{s},\rho)$ stands for the sum of hadron self-energies 
in channel $n$. Of particular interest is the meson ($h$) self-energy 
$\Pi^{(hN)}_h=(E_N/\sqrt{s})\Pi_h$ in the diagonal $n\equiv (hN)$ channel, 
where the lab self-energy $\Pi_h$ is given by   
\begin{equation} 
\Pi_h(\sqrt{s},\rho)\equiv 2\omega_hV_h=-\frac{\sqrt{s}}{E_N}4\pi
F_{hN}(\sqrt{s},\rho)\rho \; ,
\label{eq:Pi} 
\end{equation} 
depending implicitly on $\omega_h=m_h-B_h$ and on the off-shell two-body 
momenta $k,k'$. This self-energy, once evaluated {\it self-consistently} 
while converting its $\sqrt{s}$ dependence into a full density dependence, 
serves as input to the Klein-Gordon bound-state equation 
\begin{equation} 
[\:\nabla^2+{\tilde\omega}_h^2-m_h^2-\Pi_h(\omega_h,\rho)\:]\:\psi=0  
\; , 
\label{eq:KG} 
\end{equation} 
in which ${\tilde\omega}_h=\omega_h-{\rm i}\Gamma_h/2$, with $B_h$ and 
$\Gamma_h$ the binding energy and the width of the meson-nuclear bound state, 
respectively.

\begin{figure*}[htb] 
\begin{minipage}[h]{6.5cm} 
\includegraphics[width=6.0cm,height=4.6cm]{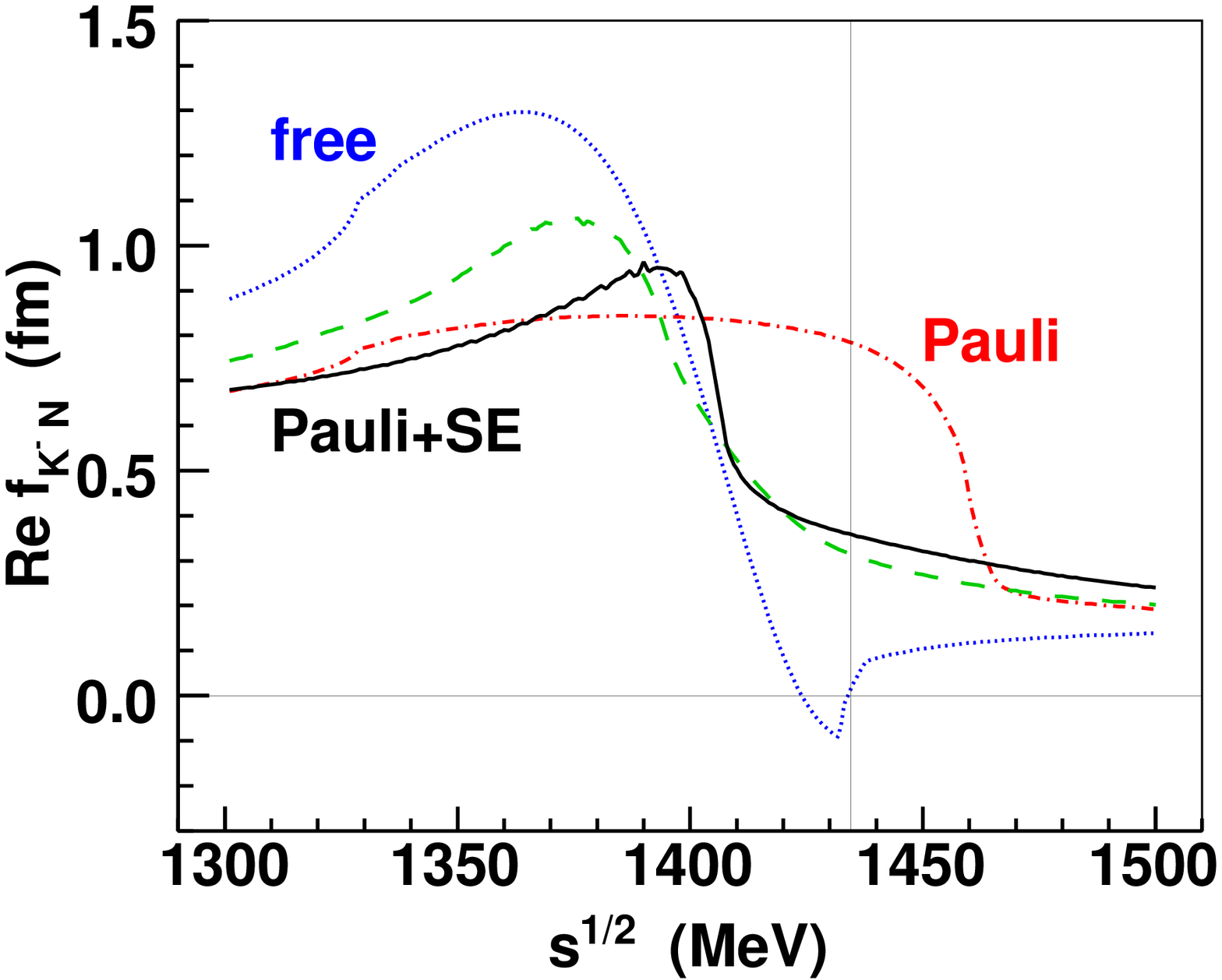}  
\includegraphics[width=6.0cm,height=4.6cm]{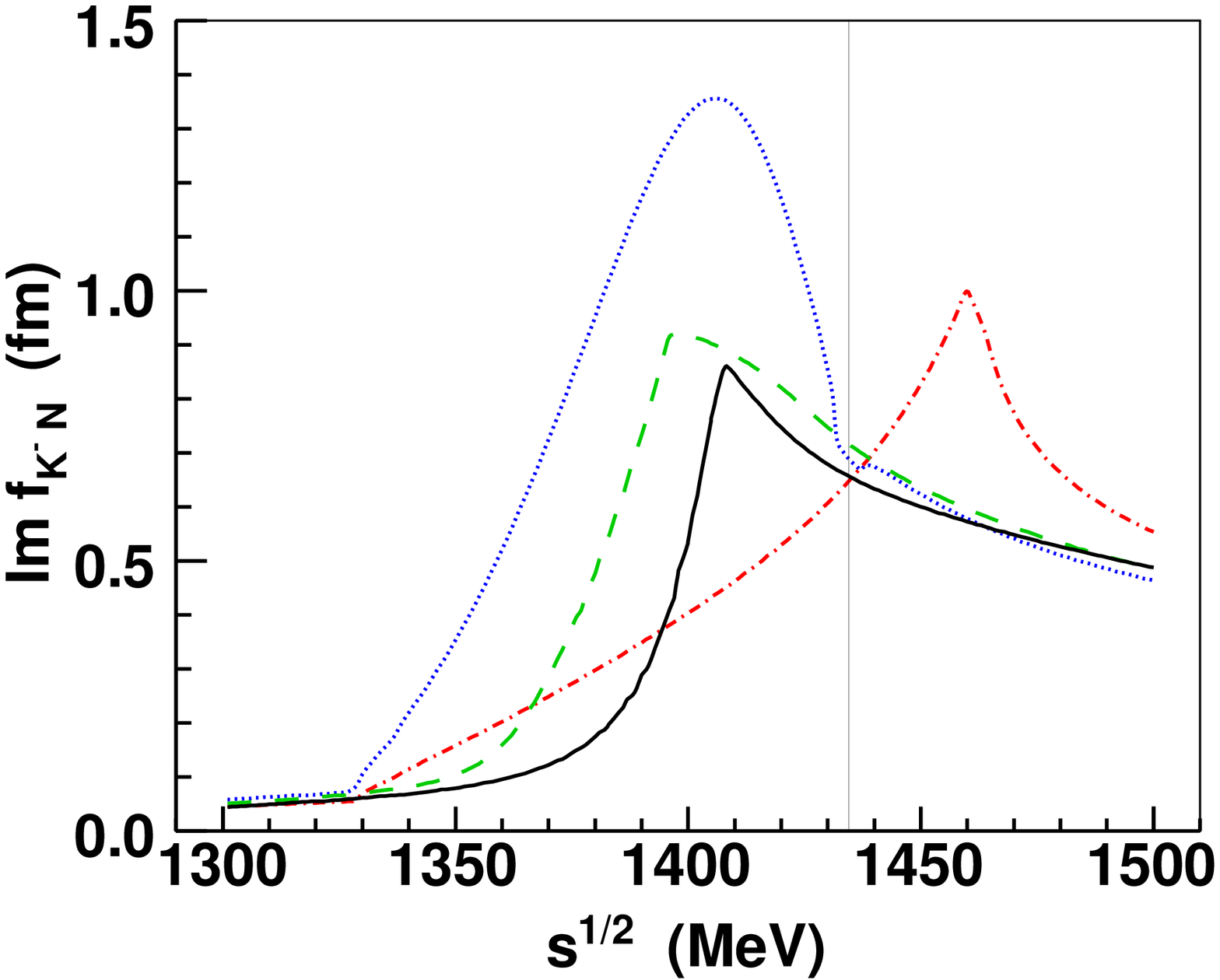} 
\end{minipage} 
\begin{minipage}[r]{6.5cm}
\includegraphics[width=6.0cm,height=4.6cm]{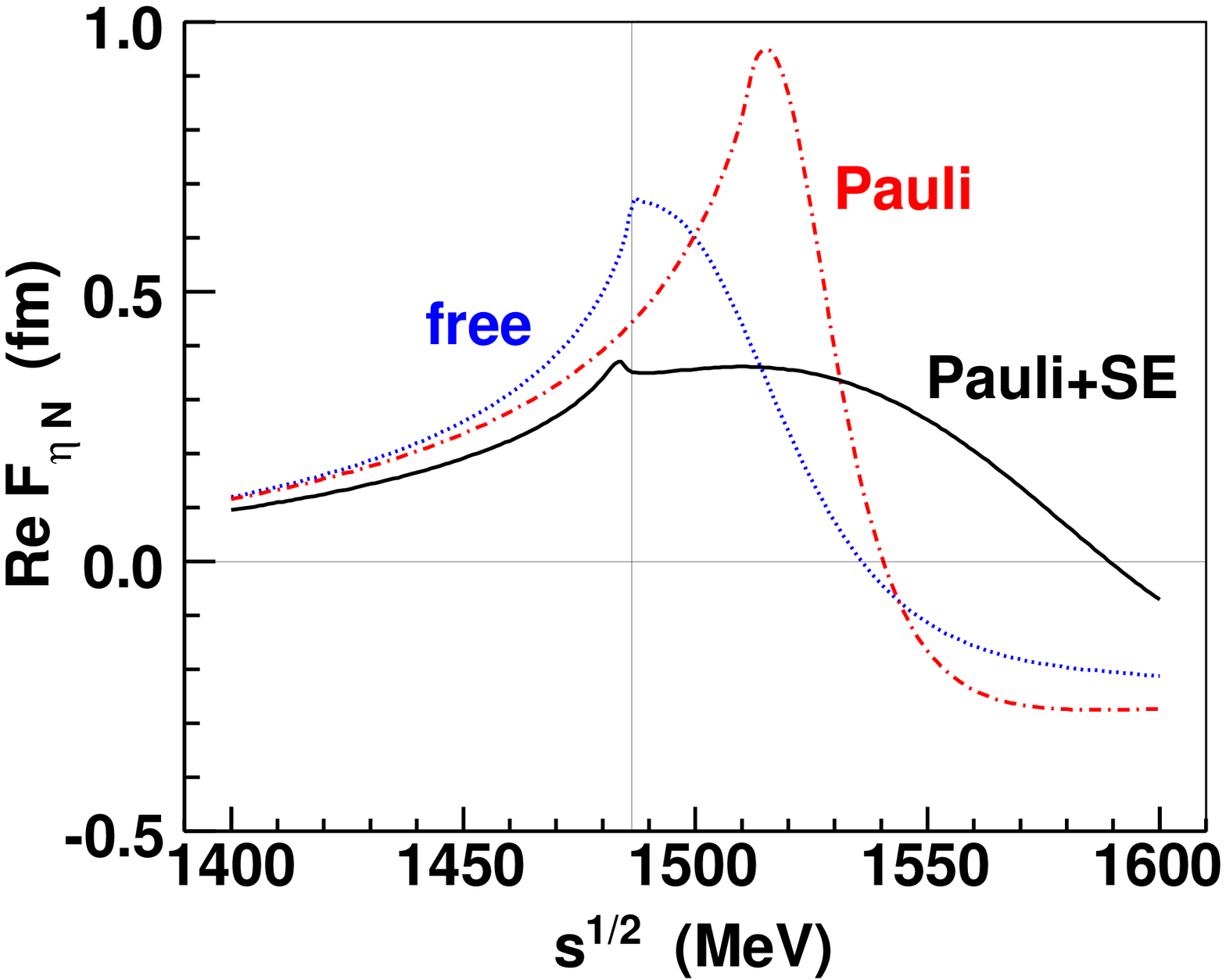} 
\includegraphics[width=6.0cm,height=4.6cm]{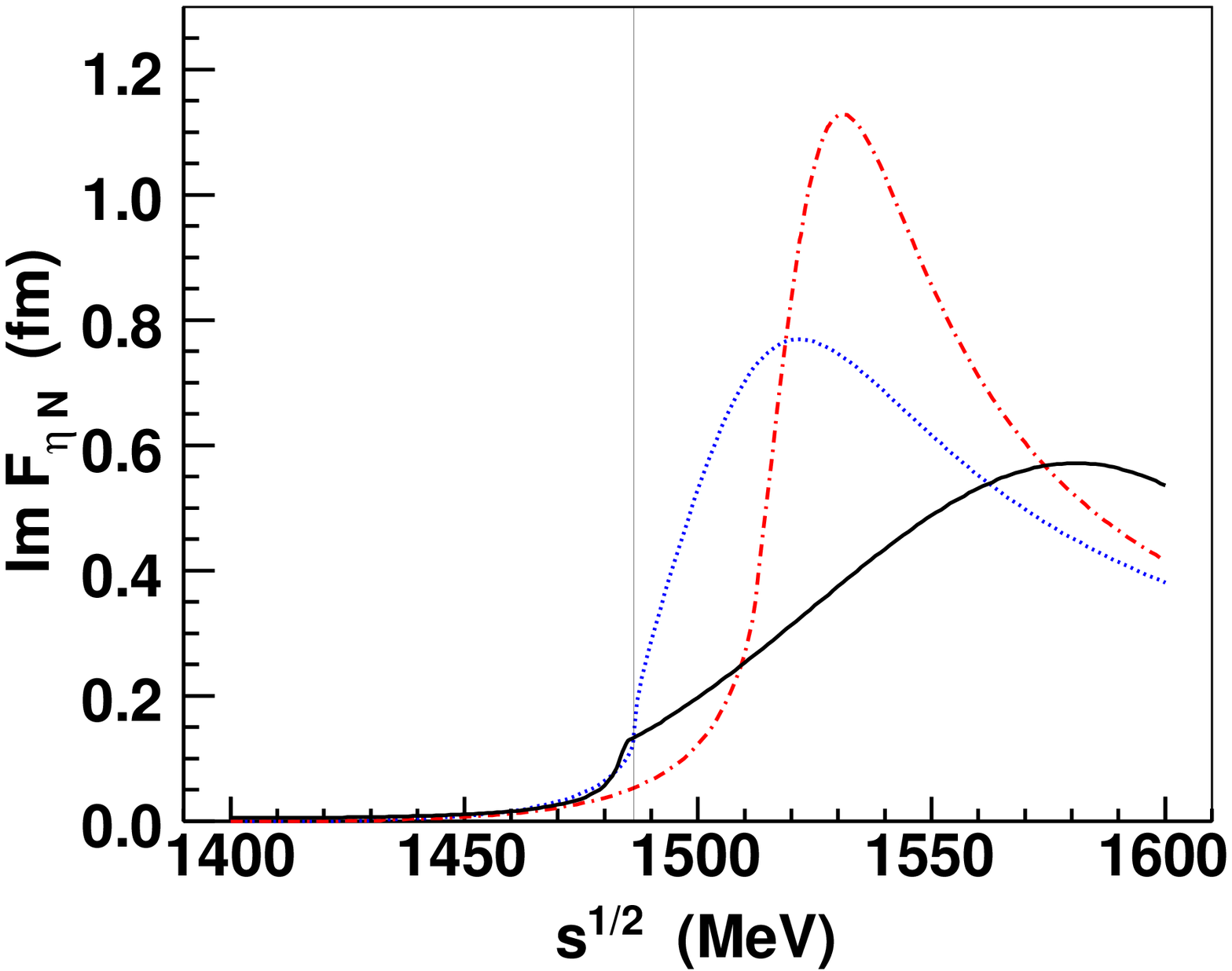} 
\end{minipage} 
\caption{Left: near-threshold energy dependence of $K^-N$ cm reduced 
scattering amplitudes in model NLO30 of Ref.~\cite{CS12} for free-space 
(dotted) and Pauli-blocked amplitudes at $\rho=\rho_0$ with (solid) and 
without (dot-dashed) meson and baryon self-energies (SE). The dashed curves 
show Pauli-blocked amplitudes with SE at $\rho=0.5\rho_0$. 
Right: energy dependence of $\eta N$ free-space and in-medium (at $\rho_0$) 
cm scattering amplitudes across threshold and beyond the $N^{\ast}$(1535) 
resonance in model NLO30$_{\eta}$ of Ref.~\cite{CS13}. The $K^-N$ and 
$\eta N$ thresholds are marked by thin vertical lines.}  
\label{fig:NLO30} 
\end{figure*}

In-medium amplitudes above and below threshold are shown in 
Fig.~\ref{fig:NLO30} for $K^-N$ \cite{CS12} and $\eta N$ \cite{CS13}. 
The $K^-N$ real part of the amplitude is strongly attractive, of order 1 fm 
at subthreshold energies that according to the discussion below are relevant 
to $K^-$ atomic and nuclear states. The attraction as well as the absorptivity 
expressed by the imaginary part of the amplitude get moderately weaker for 
$\rho\geq 0.5\rho_0$, as demonstrated by comparing on the left panel the 
solid curves ($\rho=\rho_0$) with the dashed curves ($\rho=0.5\rho_0$). 
In contrast, the $\eta N$ real part decreases substantially upon going 
below subthreshold, with values in the range of 0.1--0.2 fm, with little 
density dependence. This implies that $K^-$ bound states are very likely to 
exist, whereas $\eta$ nuclear states may not bind. Similarly, the widths 
generated by the imaginary part of the scattering amplitudes are considerably 
larger for $K^-$ than for $\eta$ mesons. 

\begin{figure}[htb]
\begin{center}
\includegraphics[width=0.48\textwidth,height=5.0cm]{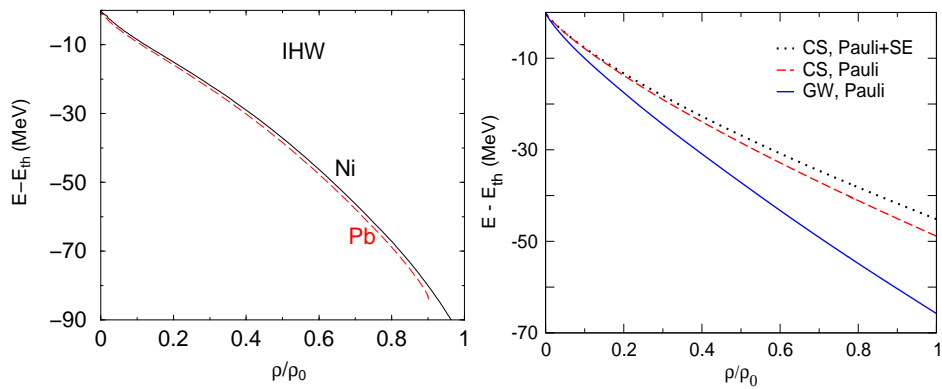} 
\includegraphics[width=0.48\textwidth,height=5.0cm]{srhocs3.eps}
\caption{Subthreshold energies probed in kaonic atoms of Ni and Pb
(left panel) and for a $1s_{\eta}$ bound state in Ca (right panel)
as a function of nuclear density, calculated self-consistently within
the IHW-based global fit to kaonic atoms \cite{FG13} and for in-medium 
$\eta N$ scattering amplitudes constructed from the meson-baryon models 
GW and CS in Ref.~\cite{CFGM13}.}
\label{fig:Evsrho}
\end{center}
\end{figure}

To determine the subthreshold energies for use in in-medium hadron-nucleon 
scattering amplitudes, we recall that the Mandelstam variable $s$ is given 
by $s=(\sqrt{s_{\rm th}}-B_{h}-B_N)^2-({\vec p}_{h}+{\vec p}_N)^2$, where 
$\sqrt{s_{\rm th}}\equiv m_{h}+m_N$ and $B_h$ and $B_N$ are meson and nucleon 
binding energies. Since ${\vec p}_{h}+{\vec p}_N\neq 0$ in the meson-nuclear 
cm frame (approximately the lab frame), the associated negative contribution 
to $s$ has to be included. To leading order in binding energies and 
kinetic energies with respect to rest masses, the downward energy shift 
$\delta\sqrt{s}\equiv\sqrt{s}-\sqrt{s_{\rm th}}$ is expressed as 
\begin{equation} 
\delta\sqrt{s}\approx - B_N - B_h - \xi_N\frac{p_N^2}{2m_N} 
- \xi_h\frac{p_h^2}{2m_h} \; ,
\label{eq:approx}
\end{equation}
where $\xi_{N(h)}\equiv m_{N(h)}/(m_N+m_h)$. Using the Fermi Gas model for 
nucleons and the local density approximation, one gets 
\begin{equation} 
\delta\sqrt{s}\approx -B_N\frac{\rho}{{\bar\rho}}-
\xi_N B_h\frac{\rho}{\rho_0}-\xi_N T_N(\frac{\rho}{\rho_0})^{2/3}-
\xi_h \frac{\sqrt{s}}{\omega_h E_N}2\pi{\rm Re}~F_{hN}(\sqrt{s},\rho)\rho \; ,
\label{eq:sqrts}
\end{equation}
where $T_N=23.0$ MeV at nuclear-matter density $\rho_0$, $B_N\approx 8.5$~MeV 
is an average nucleon binding energy and $\bar\rho$ is the average nuclear 
density. For the charged $K^-$ meson, a Coulomb term proportional to $V_C
\times(\rho/\rho_0)^{1/3}$ was added \cite{FG13}. Expression (\ref{eq:sqrts}) 
respects the low-density limit, $\delta\sqrt{s}\to 0$ upon $\rho\to 0$. For 
attractive scattering amplitudes, all four terms in Eq.~(\ref{eq:sqrts}) are 
negative definite, the last one providing substantial downward energy shift 
overlooked by many previous calculations that assumed ${\vec p}_h=0$ which 
is inappropriate for {\it finite} nuclei. Since $\sqrt{s}$ depends through 
Eq.~(\ref{eq:sqrts}) on ${\rm Re}~F_{hN}(\sqrt{s},\rho)$ which by itself 
depends on $\sqrt{s}$, it is clear that for a given value of $B_h$, $F_{hN}
(\sqrt{s},\rho)$ has to be determined {\it self-consistently} by iterating 
Eq.~(\ref{eq:sqrts}). This is done at each radial point where $\rho$ is given, 
and for each $B_h$ value during the calculation of bound states. The emerging 
correlation between the downward energy shift $\delta\sqrt{s}$ and the density 
$\rho$ renders $F_{hN}(\sqrt{s},\rho)$ into a state-dependent function of the 
density $\rho$ alone, denoted for brevity by $F_{hN}(\rho)$. This correlation 
is shown on the left panel of Fig.~\ref{fig:Evsrho} for kaonic atoms, where 
$B_{K^-}\approx 0$, and on the right panel for the $1s_{\eta}$ nuclear bound 
state in Ca. The figure demonstrates appreciable energy shifts below threshold 
in both kaonic atoms and $\eta$-nuclear bound states, amounting to 40~MeV at 
0.5$\rho_0$. 

\section{$K^-$ interactions in kaonic atoms} 
\label{sec3} 

The most recent kaonic-atom calculations are due to Friedman and Gal in 
Ref.~\cite{FG12}, using in-medium $K^-N$ scattering amplitudes generated in 
model NLO30 of Ciepl\'{y} and Smejkal \cite{CS12} as described in the previous 
section, and in Ref.~\cite{FG13} using Pauli blocked $K^-N$ scattering 
amplitudes generated from the free-space NLO scattering amplitude of Ikeda, 
Hyodo and Weise \cite{IHW11}. The CS \cite{CS12} and IHW \cite{IHW11} 
free-space amplitudes $F_{K^-N}(\sqrt{s})$ agree semi-quantitatively with each 
other. The kaonic-atom fit in Ref.~\cite{FG13} considers in addition to the 
input in-medium IHW-based one-nucleon (1N) amplitude $F_{K^-N}(\sqrt{s},\rho)$ 
also many-nucleon absorptive and dispersive contributions, represented by 
energy-independent phenomenological amplitude $F^{\rm many}_{K^-N}(\rho)$ 
with prescribed density dependence form that includes several fit parameters. 
The assumption of energy independence is motivated by observing that $K^-$ 
absorption on two nucleons, which is expected to dominate 
$F^{\rm many}_{K^-N}$, releases energy $\sim$$m_{K^-}$ considerably larger 
than the subthreshold energies of less than 100 MeV encountered in kaonic-atom 
calculations. The self-energy input $\Pi_{K^-}$ to the KG equation 
(\ref{eq:KG}) is now constructed from an {\it effective} $K^-N$ scattering 
amplitude $F^{\rm eff}_{K^-N}=F^{\rm one}_{K^-N}+F^{\rm many}_{K^-N}$ which 
is iterated through the self-consistency expression (\ref{eq:sqrts}). 
This introduces coupling between the many-nucleon fitted amplitude 
$F^{\rm many}_{K^-N}$ and the converged one-nucleon amplitude 
$F^{\rm one}_{K^-N}$ that evolves from the 1N input amplitude 
$F_{K^-N}(\rho)$: $F^{\rm one}_{K^-N}(\rho)\to F_{K^-N}(\rho)$ upon 
$F^{\rm many}_{K^-N}\to 0$. 

\begin{figure}[htb] 
\begin{center} 
\includegraphics[width=0.50\textwidth]{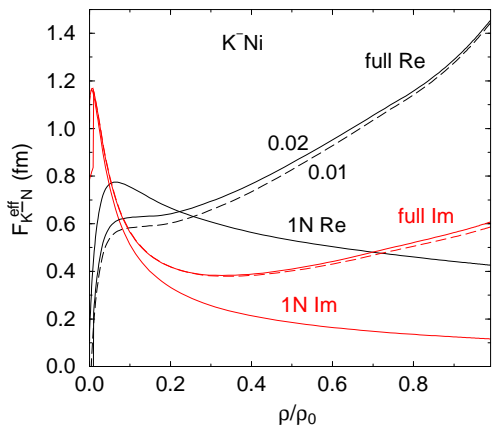} 
\caption{Kaonic-atom globally fitted amplitude $F^{\rm eff}_{K^-}(\rho)$, 
marked ``full", and the in-medium IHW-based amplitude $F_{K^-N}(\rho)$ in 
the absence of many-nucleon contributions, marked ``1N", as a function of 
nuclear density in Ni. Solid (dashed) curves are for matching to free-space 
amplitudes at 0.02(0.01)$\rho_0$.} 
\label{fig:IHWampl} 
\end{center} 
\end{figure} 

The full effective amplitude $F^{\rm eff}_{K^-N}(\rho)$ resulting from the 
global kaonic-atom fit in \cite{FG13} is shown in Fig.~\ref{fig:IHWampl} 
marked ``full", along with the in-medium IHW-based amplitude $F_{K^-N}(\rho)$ 
marked ``1N". The figure makes it clear that for densities exceeding 
$\sim$0.5$\rho_0$ the full effective amplitude departs appreciably from the 
in-medium IHW-based amplitude, which in the case of the imaginary part amounts 
to doubling the 1N absorptivity of in-medium $K^-$ mesons. For a more details 
we refer the reader to Ref.~\cite{FG13}. 

\begin{figure}[htb] 
\begin{center} 
\includegraphics[width=0.48\textwidth,height=7.6cm]{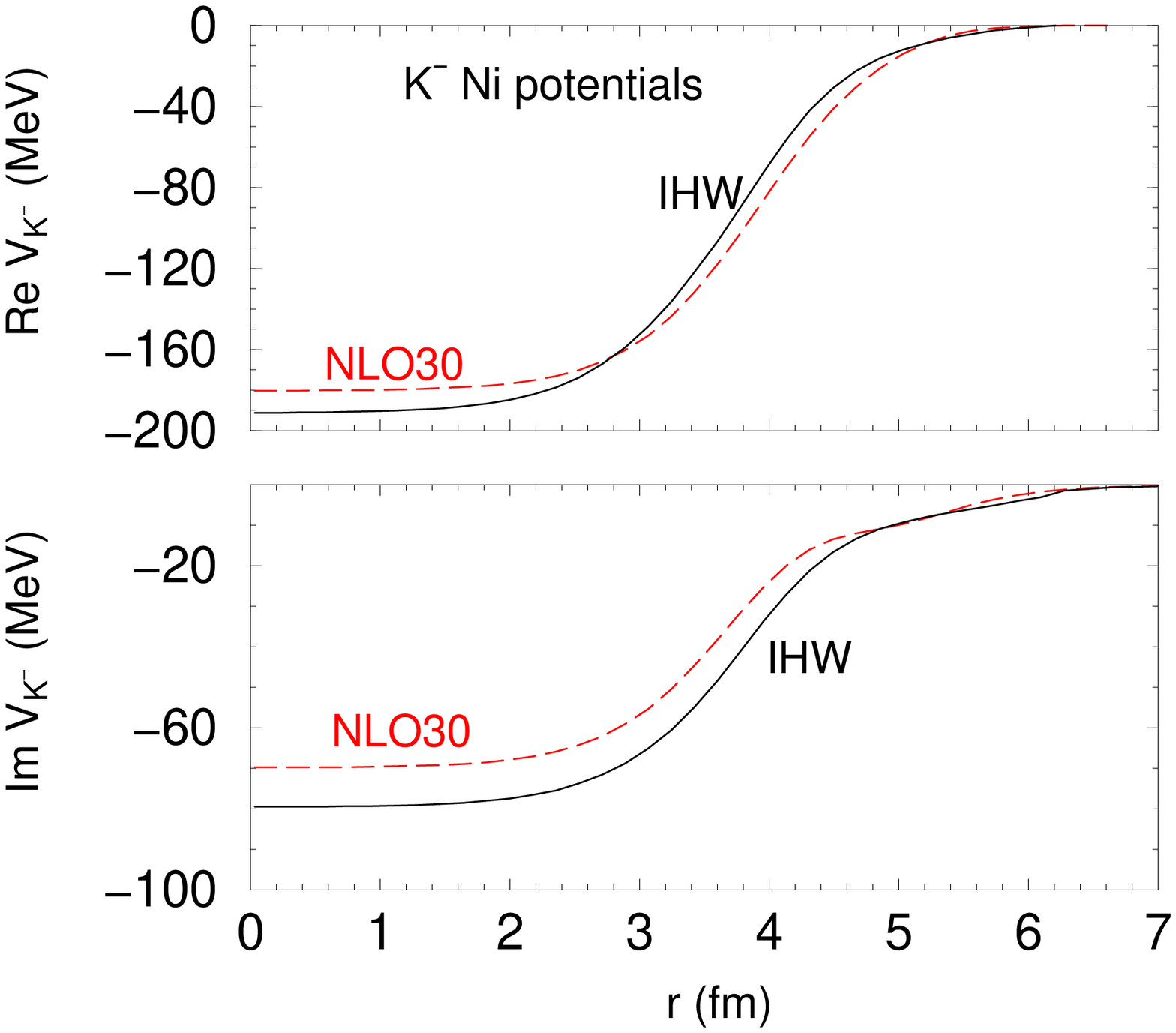} 
\includegraphics[width=0.48\textwidth,height=7.6cm]{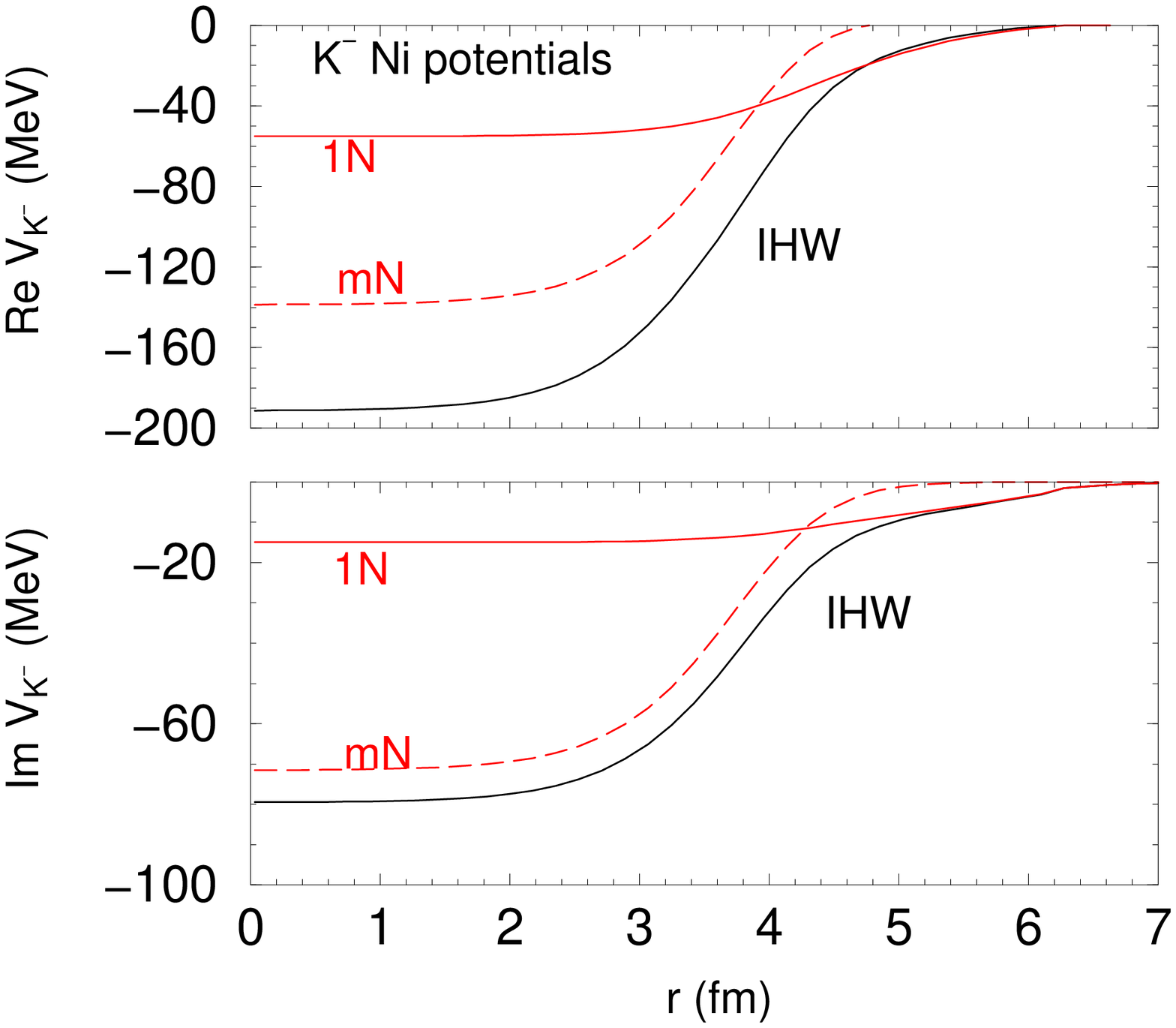} 
\caption{Left: $K^-$ nuclear potentials for $K^-$ atoms of Ni derived 
from global fits based on in-medium IHW amplitudes \cite{FG13}, with the 
corresponding $1N$ and many-nucleon ($mN$) components on the right panel. 
The dashed curves in the left panel are derived from in-medium NLO30 
amplitudes \cite{FG12}. The rms radius of the input Ni density is 3.72~fm.} 
\label{fig:Kpot} 
\end{center} 
\end{figure}

The $K^-$ nuclear attraction and absorptivity deduced from global kaonic-atom 
fits are sizable at central nuclear densities. This is demonstrated by the 
real and imaginary parts of the potential $V_{K^-}$ plotted for Ni in 
Fig.~\ref{fig:Kpot}. Although the potential depths might reflect a smooth 
extrapolation provided by the input components of the $K^-N$ amplitudes, the 
potential at 0.5$\rho_0$ and perhaps up to 0.9$\rho_0$ is reliably determined 
in kaonic-atom fits \cite{BF07}. It is reassuring that both IHW-based and 
NLO30-based fits agree with each other semi-quantitatively as shown on the 
left panel. 

The right panel of Fig.~\ref{fig:Kpot} shows a non-additive splitting of the 
fitted $K^-$-nuclear potential into a 1N in-medium component, obtained on 
the assumption that there is no many-nucleon (mN) component present, and a 
fitted mN component. The breakdown of the imaginary part of the potential is 
of particular interest, indicating that the mN component which is sizable in 
the nuclear interior becomes negligible about half a fermi outside of the 
half-density radius. This has implications for choosing optimally kaonic-atom 
candidates where widths of two atomic levels can be measured so as to 
substantiate the 1N vs mN pattern observed in global fits \cite{FO13}.

\section{Few-body kaonic quasibound states}
\label{sec4}

\begin{figure}[b!] 
\begin{center} 
\includegraphics[width=0.90\textwidth]{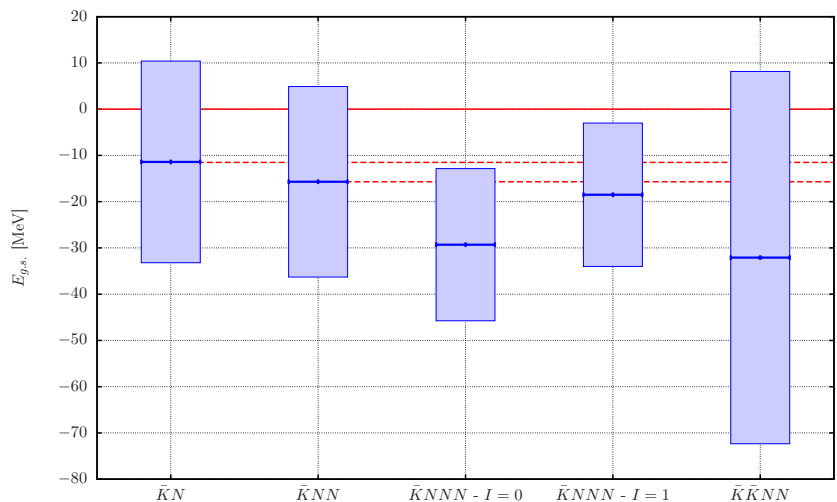} 
\caption{Calculated binding energies and $\bar K N \to \pi Y$ widths of 
$\bar K$ and $\bar K\bar K$ few-body quasibound states \cite{BGL12} in MeV. 
Horizontal lines denote particle-stability thresholds. Widths are 
represented by vertical bars.} 
\label{fig:BGL} 
\end{center} 
\end{figure}

For $K^-$-nuclear three- and four-body calculations, a variant of the downward 
energy shift Eq.~(\ref{eq:sqrts}) derived for many-body calculations was 
formulated by Barnea, Gal and Liverts \cite{BGL12}: 
\begin{equation} 
\delta\sqrt{s} = -\frac{B}{A}-\frac{A-1}{A}B_K-
\xi_{N}\frac{A-1}{A}\langle T_{N:N} \rangle -\xi_{K}\left ( \frac{A-1}{A} 
\right )^2 \langle T_K \rangle \; ,
\label{eq:sqrt{s}} 
\end{equation} 
with $A$ the baryonic number, $B$ the total binding energy of the system, 
$B_K=-E_K$, $T_K$ the kaon kinetic energy operator in the total cm frame 
and $T_{N:N}$ the pairwise $NN$ kinetic energy operator in the $NN$ pair 
cm system. Note that $\delta\sqrt{s}$ is negative-definite by expression 
(\ref{eq:sqrt{s}}) which provides a self-consistency cycle upon requiring that 
$\sqrt{s}$ derived through Eq.~(\ref{eq:sqrt{s}}) from the solution of the 
Schroedinger equation agrees with the value of $\sqrt{s}$ used for the input 
$V_{\bar K N}(\sqrt{s})$. Total binding energies calculated variationally in 
the hyperspherical basis are shown in Fig.~\ref{fig:BGL} for three- 
and four-body kaonic bound states. Details of the input chirally-based 
energy-dependent $\bar K N$ interactions and the actual calculations of 
these few-body kaonic clusters are given in Ref.~\cite{BGL12}.  

\newpage
Shown also as vertical bars in Fig.~\ref{fig:BGL} are $\bar K N \to \pi Y$ 
width estimates using the approximation 
\begin{equation} 
\frac{\Gamma}{2}\approx\langle \,\Psi_{\rm g.s.} |
-{\rm Im}\,{\cal V}_{\bar KN}\, | \, \Psi_{\rm g.s.} \, \rangle \;, 
\label{eq:Gamma} 
\end{equation} 
where ${\cal V}_{\bar KN}$ consists of all pairwise $\bar KN$ interactions. 
Expression (\ref{eq:Gamma}) provides a good approximation owing to $|{\rm Im}
\,{\cal V}_{\bar KN}|\ll |{\rm Re}\,{\cal V}_{\bar KN}|$ \cite{HW08}. 
Expressions similar to (\ref{eq:sqrt{s}}) and (\ref{eq:Gamma}) were used 
in ${\bar K}{\bar K}NN$ calculations. With $\bar K N$ input interactions 
that become weaker upon going subthreshold \cite{HW08}, and owing to the 
self-consistency requirement, the calculated binding energies (widths) come 
out typically 10 (10--40) MeV lower than for threshold input interactions 
$V_{\bar KN}(\sqrt{s_{\rm th}})$. In particular, the $I=1/2$ $\bar KNN$ g.s., 
known as `$K^-pp$', lies only 4.3 MeV below the 11.4 MeV centroid of the $I=0$ 
$\bar KN$ quasibound state, the latter value differing substantially from the 
27 MeV binding energy assigned traditionally to the $\Lambda(1405)$ resonance 
(and used in non-chiral calculations). The widths exhibited in the figure, of 
order 40 MeV for single-$\bar K$ clusters and twice that for double-$\bar K$ 
clusters, are due to $\bar K N \to \pi Y$. Additional $\bar K NN\to YN$ 
contributions of up to $\sim$10~MeV in $K^-pp$ \cite{DHW08} and $\sim$20~MeV 
in the four-body systems \cite{BGL12} are foreseen.

\section{Many-body kaonic quasibound states} 
\label{sec5} 

\begin{figure}[htb]
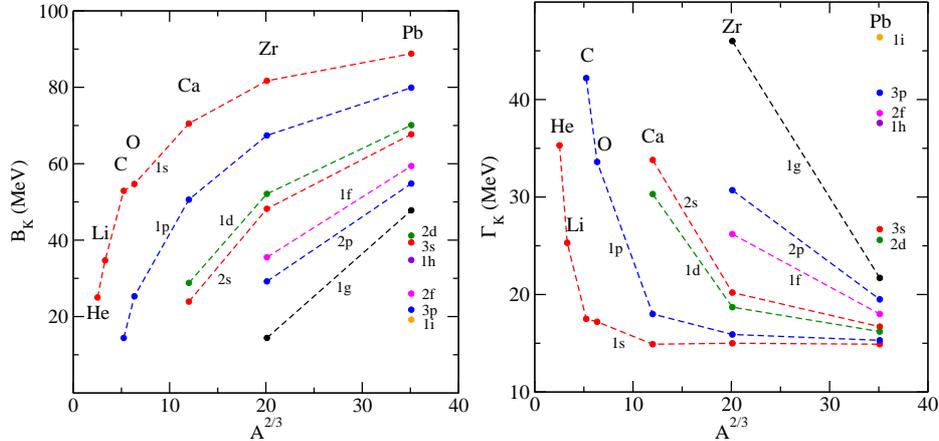
 
\begin{center} 
\includegraphics[width=0.48\textwidth]{gm12_fig6a.eps} 
\includegraphics[width=0.48\textwidth]{gm12_fig6b.eps} 
\caption{$K^-$ Binding energies $B_K$ (left) and widths $\Gamma_K$ (right) 
calculated self-consistently using NLO30-based in-medium $K^-N$ subthreshold 
amplitudes and static RMF densities \cite{GM12}.}
\label{fig:GM} 
\end{center} 
\end{figure}

In-medium $\bar KN$ scattering amplitudes derived from the chirally motivated 
NLO30 model by CS \cite{CS12} were used by Gazda and Mare\v{s} \cite{GM12} 
to evaluate $K^-$ quasibound states across the periodic table, with binding 
energies and widths shown in Fig.~\ref{fig:GM}. The phenomenological mN 
amplitudes derived from fitting kaonic-atom data, as described in 
Sect.~\ref{sec3}, are not included in these calculations. The left panel 
demonstrates a robust pattern of $K^-$-nuclear binding owing to the strongly 
attractive in-medium $K^-N$ scattering amplitude shown in Fig.~\ref{fig:NLO30}, 
in agreement with previous calculations by Weise and collaborators who used 
a less advanced form of the $\delta\sqrt{s}$ self-consistency requirement 
\cite{WH08}. Several $K^-$ quasibound states are predicted to exist in 
a given nucleus, except for the very light nuclei, but as suggested by the 
corresponding widths shown on the right panel these states are likely to be 
too broad to be uniquely resolved. 

The hierarchy of widths shown for a given nucleus in Fig.~\ref{fig:GM} is 
also worth noting. For energy independent potentials one expects maximal 
widths in the lowest, most localized $1s_K$ states, and gradualy decreased 
widths in excited states which are less localized within the nucleus. The 
reverse is observed on the right panel of the figure. This is a corollary of 
requiring self consistency: the more excited a $K^-$ quasibound state is, the 
lower nuclear density it feels, and a smaller downward shift into subthreshold 
energies it probes via the $s(\rho)$ dependence. Since by Fig.~\ref{fig:NLO30} 
Im$\;f_{K^-N}(\rho)$ decreases strongly upon going below threshold, its 
contribution to the calculated width gets larger, the higher the excited 
quasibound-state energy is. Additional width contributions from mN processes 
are found to increase appreciably the calculated widths, with estimates made 
in Ref.~\cite{GM12} for the overall width of $K^-$ quasibound states in Ca, 
for example, in the range of values $\Gamma_K\sim (50-70)$~MeV.

\section{$\eta$-nuclear quasibound states} 
\label{sec6} 

\begin{figure}[htb]
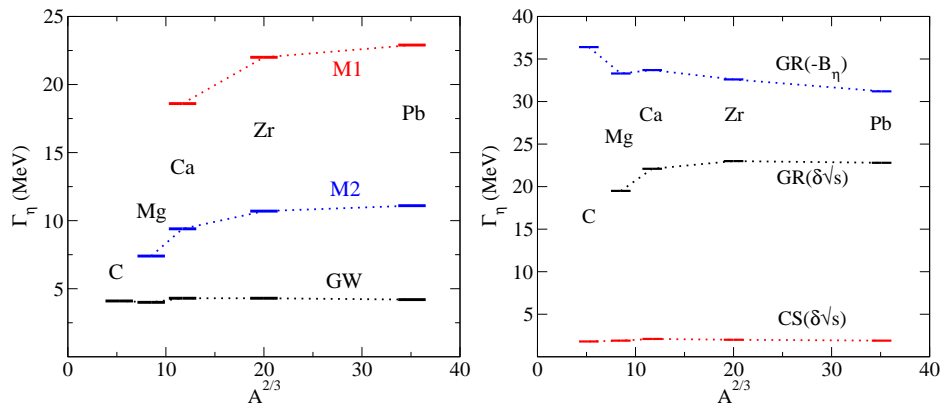
 
\begin{center} 
\includegraphics[width=0.48\textwidth]{gamma-se.eps} 
\includegraphics[width=0.48\textwidth]{gamma-ales-oset.eps} 
\caption{Calculated widths of $1s_{\eta}$ nuclear bound states are shown 
on the left panel for in-medium GW \cite{GW05} and M \cite{MBM12} versions 
of $\eta N$ scattering amplitudes, and on the right panel for the in-medium 
versions CS \cite{CFGM13} and GR \cite{GR02}, the latter is based on the 
free-space model IOV \cite{IOV02} and its in-medium extension \cite{IO02}.} 
\label{fig:gamma} 
\end{center} 
\end{figure}

It was noted in Sect.~\ref{sec1} that free-space near-threshold $\eta N$ 
scattering amplitudes $F_{\eta N}(\sqrt{s})$ are highly model dependent, 
as demonstrated in Fig.~\ref{fig:aEtaN1} for four of the many amplitudes 
available in the literature. Since the amplitudes denoted there GW \cite{GW05} 
and M2 (also M1) \cite{MBM12} are available only in free-space versions, 
appropriate in-medium versions have been produced in Ref.~\cite{FGM13} by 
applying the Ericson-Ericson multiple-scattering renormalization \cite{WRW97}: 
\begin{equation}
F_{\eta N}(\sqrt{s},\rho)=\frac{F_{\eta N}(\sqrt{s})}
{1+\xi(\rho)(\sqrt{s}/m_N)F_{\eta N}(\sqrt{s})\rho}\;,
\label{eq:WRW}
\end{equation}
where $p_F=(3{\pi}^2\rho/2)^{1/3}$ is the local Fermi momentum corresponding 
to density $\rho$ and $\xi(\rho)=9\pi/4p_F^2$ accounts for Pauli blocking 
at threshold. These in-medium amplitudes were then used as input within 
self-consistent calculations, as described in Sect.~\ref{sec2}. 
In the left panel of Fig.~\ref{fig:gamma} we show $1s_{\eta}$-nuclear widths 
calculated in these models. Whereas model GW generates widths smaller than 
5 MeV, the widths in models M are twice (M2) and five times (M1) as large. 
This demonstrates clearly that the calculated $\eta$-nuclear widths are not 
related directly to Im$\;a_{\eta N}$, the imaginary part of the scattering 
length which is approximately the same in all these models. The difference 
between the calculated sequences of widths is in fact due to the difference 
between the subthreshold free-space imaginary parts of the amplitudes shown 
in the right panel of Fig.~\ref{fig:aEtaN1}, in spite of their near equality 
at threshold. 

In-medium versions that account for Pauli blocking and self-energies 
are available for the other two model amplitudes CS \cite{CS13} and IOV 
\cite{IOV02} shown in Fig.~\ref{fig:aEtaN1}. These in-medium versions, 
marked CS and GR \cite{GR02} respectively in Fig.~\ref{fig:gamma}, 
have been discussed extensively in Ref.~\cite{CFGM13} with calculated 
$1s_{\eta}$-nuclear widths shown on the right panel of the figure. 
In this plot, calculations using the self-consistency requirement 
(\ref{eq:sqrts}) are denoted $\delta\sqrt{s}$ and the GR calculations 
that used a density-independent $\delta\sqrt{s}=-B_{\eta}$ self-consistency 
requirement are denoted $-B_{\eta}$. The calculated GR widths are considerably 
larger than those due to CS, which again results from the difference between 
the subthreshold free-space imaginary parts of the amplitudes shown on the 
right panel of Fig.~\ref{fig:aEtaN1}, in spite of their near equality at 
threshold. It is also seen that applying the $\delta\sqrt{s}$ procedure 
instead of their original $-B_{\eta}$ procedure leads to appreciable 
reduction of the calculated GR widths. The main conclusion drawn from 
Fig.~\ref{fig:gamma} is that, while models GW and CS produce sufficiently 
small $\eta$-nuclear widths, it will be prohibitively difficult to resolve 
$\eta$-nuclear states if the correct underlying $\eta N$ amplitudes are 
due to models M1, M2 or GR. The widths presented in Fig.~\ref{fig:gamma} 
do not include contributions from two-nucleon processes which are estimated 
to add a few MeV \cite{wycech10}. 

\begin{figure}[htb]
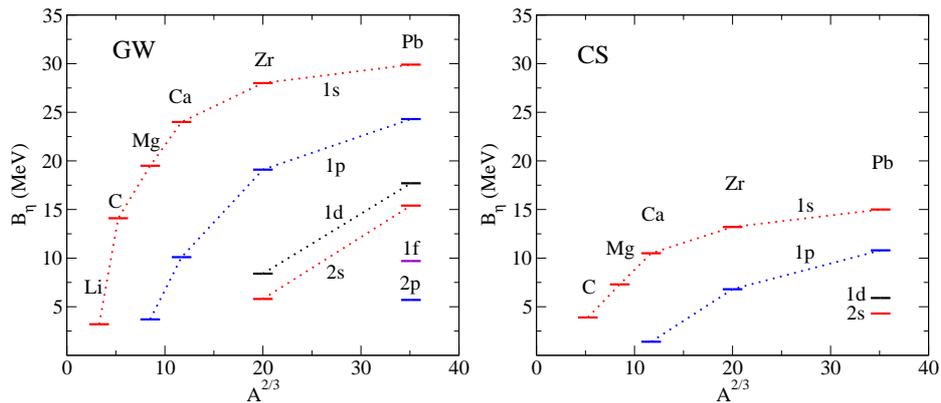
 
\begin{center} 
\includegraphics[width=0.48\textwidth]{beta2.eps} 
\includegraphics[width=0.48\textwidth]{beta-ales+se2.eps} 
\caption{Spectra of $\eta$-nuclear single-particle bound states across the 
periodic table, calculated self-consistently using in-medium models of the 
$\eta N$ subthreshold scattering amplitude, are shown in the left panel for 
the GW model \cite{GW05} and in the right panel for the NLO30$_{\eta}$ model 
of CS \cite{CS13}. Pauli blocking is included for both in-medium models, 
whereas hadron self-energies are accounted for only in the CS-based 
calculations.} 
\label{fig:GW-CS} 
\end{center} 
\end{figure} 

Finally, in Fig.~\ref{fig:GW-CS} we show $\eta$-nuclear single-particle 
spectra across the periodic table, evaluated for models GW and CS in which 
the calculated widths turned out to be sufficiently small to resolve 
individual bound states. Both in-medium versions of these models account for 
Pauli blocking, whereas CS also accounts for hadron self-energies, resulting 
in 2--3 MeV lower binding energies relative to those calculated with only 
Pauli blocking. 

\section{Summary and outlook} 
\label{sec7} 

In this overview of $\bar K$- and $\eta$-nuclear bound-state calculations we 
have focused on the role played by the underlying meson-baryon subthreshold 
dynamics. It was shown how the energy dependence of the meson-baryon 
in-medium scattering amplitudes is converted into density dependence 
of the meson self-energies, or equivalently of meson-nucleus optical 
potentials. Based on global fits of $K^-$-atom data we argued that 
in-medium chiral model input has to be supplemented by appreciable 
many-nucleon dispersive and absorptive potential contributions which imply 
uniformly large widths of order 50 MeV and more for $\bar K$-nuclear bound 
states, except perhaps for the very light few-body systems. Smaller widths, 
of order 20 MeV or less, were calculated for $\eta$-nuclear bound states. 
This will make it difficult to identify uniquely such states in forthcoming 
experiments, unless the underlying $\eta N$ physics corresponds to models such 
as GW or CS. Furthermore, the in-medium subthreshold amplitudes encountered 
in $\eta$-nuclear bound-state calculations are substantially weaker both 
in their real part as well as in their imaginary part than the $\eta N$ 
scattering length. This weakening of the real part makes the binding of 
$\eta$ in very light nuclei such as $^3$He and $^4$He improbable, except 
perhaps in model GW. However, the methodology of constructing and using 
$\eta$-nuclear potentials does not fit into realistic few-body calculations 
which require separate treatment. To date, in spite of several experimental 
searches for $\eta$-nuclear bound states, particularly in the He isotopes 
(for the most recent report see Ref.~\cite{COSY13}), the only claim of 
observing such a bound state is in the reaction $p+{^{27}{\rm Al}}\to 
{^3{\rm He}}+{_{\eta}^{25}{\rm Mg}}\to {^3{\rm He}}+p+\pi^-+X$, reported 
by the COSY-GEM collaboration \cite{GEM09}.

\section*{Acknowledgements} 
A.G. would like to thank Pawel Moskal for the invitation to participate in 
the Mesic Nuclei symposium and for his kind hospitality. This work was 
supported by the GACR Grant No. 203/12/2126, as well as by the EU initiative 
FP7, HadronPhysics3, under the SPHERE and LEANNIS cooperation programs.


\begin{thebibliography}{99} 

\bibitem{wycech71} S.~Wycech, Nucl. Phys. B 28 (1971) 541.  
  
\bibitem{BT72} W.A.~Bardeen, E.W.~Torigoe, Phys. Lett. B 38 (1972) 135. 

\bibitem{rook75} J.R.~Rook, Nucl. Phys. A 249 (1975) 466. 

\bibitem{CFGGM11} A.~Ciepl\'{y}, E.~Friedman, A.~Gal, D.~Gazda, J.~Mare\v{s}, 
Phys. Lett. B 702 (2011) 402, Phys. Rev. C 84 (2011) 045206. 

\bibitem{CFGK11} A.~Ciepl\'{y}, E.~Friedman, A.~Gal, V.~Krej\v{c}i\v{r}\'{i}k,
Phys. Lett. B 698 (2011) 226. 

\bibitem{BGL12} N.~Barnea, A.~Gal, E.Z.~Liverts, Phys. Lett. B 712 (2012) 132. 

\bibitem{FG12} E.~Friedman, A.~Gal, Nucl. Phys. A 881 (2012) 150. 

\bibitem{GM12} D.~Gazda, J.~Mare\v{s}, Nucl. Phys. A 881 (2012) 159. 

\bibitem{FG13} E.~Friedman, A.~Gal, Nucl. Phys. A 899 (2013) 60. 

\bibitem{FGM13} E.~Friedman, A.~Gal, J.~Mare\v{s}, Phys. Lett. B 725 (2013) 
334. 

\bibitem{CFGM13} A.~Ciepl\'{y}, E.~Friedman, A.~Gal, J.~Mare\v{s}, 
Nucl. Phys. A 925 (2014) 126. 

\bibitem{GO13} Z.-H.~Guo, J.A.~Oller, Phys. Rev. C 87 (2013) 035202. 

\bibitem{SID11} M.~Bazzi, et al., SIDDHARTA Collaboration, Phys. Lett. B 704 
(2011) 113, Nucl. Phys. A 881 (2012) 88. 

\bibitem{GW05} A.M.~Green, S.~Wycech, Phys. Rev. C 71 (2005) 014001. 

\bibitem{CS13} A.~Ciepl\'{y}, J.~Smejkal, Nucl. Phys. A 919 (2013) 46. 


\bibitem{MBM12} M.~Mai, P.C.~Bruns, U.-G.~Mei{\ss}ner, Phys. Rev. D 86 (2013) 
094033. 

\bibitem{IOV02} T.~Inoue, E.~Oset, M.J.~Vicente Vacas, Phys. Rev. C 65 (2002) 
035204. 

\bibitem{IHW11} Y.~Ikeda, T.~Hyodo, W.~Weise, Phys. Lett. B 706 (2011) 63, 
Nucl. Phys. A 881 (2012) 98. 

\bibitem{CS12} A.~Ciepl\'{y}, J.~Smejkal, Nucl. Phys. A 881 (2012) 115. 

\bibitem{BF07} N.~Barnea, E.~Friedman, Phys. Rev. C 75 (2007) 022202(R). 

\bibitem{FO13} E.~Friedman, S.~Okada, Nucl. Phys. A 915 (2013) 170. 

\bibitem{HW08} T.~Hyodo, W.~Weise, Phys. Rev. C 77 (2008) 035204. 

\bibitem{DHW08} A.~Dot\'{e}, T.~Hyodo, W.~Weise, Nucl. Phys. A 804 
(2008) 197, Phys. Rev. C 79 (2009) 014003.  

\bibitem{WH08} W.~Weise, R.~H\"{a}rtle, Nucl. Phys. A 804 (2008) 173. 

\bibitem{WRW97} T.~Waas, M.~Rho, W.~Weise, Nucl. Phys. A 617 (1997) 449. 

\bibitem{GR02} C.~Garc\'{i}a-Recio, T.~Inoue, J.~Nieves, E.~Oset, 
Phys. Lett. B 550 (2002) 47. 

\bibitem{IO02} T.~Inoue, E.~Oset, Nucl. Phys. A 710 (2002) 354. 

\bibitem{wycech10} S.~Wycech, Acta Phys. Polon. B 41 (2010) 2201. 

\bibitem{COSY13} P.~Adlarson, et al. (WASA-at-COSY Collaboration), 
Phys. Rev. C 87 (2013) 035204. 

\bibitem{GEM09} A.~Budzanowski, et al. (COSY-GEM Collaboration), 
Phys. Rev. C 79 (2009) 012201(R). 


\end{thebibliography}
\end{document}